\documentclass[structabstract]{aa}  
\def\FLASH{{\sc flash}}
\def\PARAMESH{{\sc paramesh}}
\usepackage{natbib}

\usepackage{graphicx}
\usepackage{txfonts}
\begin{document}
   \title{On the observability of T Tauri accretion shocks in the X-ray
band}

   \subtitle{}

   \author{G. G. Sacco\inst{1,}\inst{2} \and 
S. Orlando\inst{2} \and 
C. Argiroffi\inst{3,}\inst{2} \and
A. Maggio\inst{2} \and 
G. Peres\inst{3,}\inst{2} \and
F. Reale\inst{3,}\inst{2} \and
R. L. Curran\inst{2,4}
}

  \institute{Chester F. Carlson Center for Imaging Science, Rochester
Institute of Technology, 54 Lomb Memorial Dr., Rochester, NY 14623,
USA\\ \email{sacco@cis.rit.edu} \and
INAF-Osservatorio Astronomico di Palermo, Piazza del Parlamento, 1,
90134, Palermo, Italy\\ \and 
DSFA-Universit\`a degli Studi di Palermo, Piazza del Parlamento, 1,
90134, Palermo, Italy\\ \and
Department of Physics, Rochester Institute of Technology, 54 Lomb Memorial Dr., Rochester, NY 14623,
USA}

   \date{Received May 6 2010/Accepted July 13 2010}

 
 \abstract 
{High resolution X-ray observations of classical T Tauri stars
(CTTSs) show a soft X-ray excess due to high density plasma ($n_{\rm
e}=10^{11}-10^{13}$ cm$^{-3}$). This emission has been attributed to
shock-heated accreting material impacting onto the stellar surface.}
{We investigate the observability of the shock-heated accreting material
in the X-ray band as a function of the accretion stream properties
(velocity, density, and metal abundance) in the case of plasma-$\beta
\ll 1$ (thermal pressure $\ll$ magnetic pressure) in the post-shock
zone.}
{We use a 1-D hydrodynamic model describing the impact of an accretion
stream onto the chromosphere of a CTTS, including the effects of radiative
cooling, gravity stratification and thermal conduction. We explore the
space of relevant parameters and synthesize from the model results the
X-ray emission in the $[0.5-8.0]$ keV band and in the resonance lines
of \ion{O}{vii} (21.60 {\AA}) and \ion{Ne}{ix} (13.45 {\AA}), taking
into account the absorption from the chromosphere.}
{The accretion stream properties largely influence the temperature
and the stand-off height of the shocked slab and its sinking in the
chromosphere, determining the observability of the shocked plasma
affected by chromospheric absorption. Our model predicts that X-ray
observations preferentially detect emission from low density and high
velocity shocked accretion streams due to the large absorption of
dense post-shock plasma. In all the cases examined, the post-shock zone
exhibits quasi-periodic oscillations due to thermal instabilities with
periods ranging from $3\times 10^{-2}$ to $4\times 10^3$ s. In the case
of inhomogeneous streams and $\beta \ll 1$, the shock oscillations are
hardly detectable.}
{We suggest that, if accretion streams are inhomogeneous, the selection
effect introduced by the absorption on observable plasma components
may easily explain the discrepancy between the accretion rate measured
by optical and X-ray data as well as the different densities measured
using different He-like triplets in the X-ray band.}


   \maketitle
%

\section{Introduction}

Young accreting stars, also called classical T Tauri stars (CTTSs), have
been well studied during the last decades, both from the observational
and theoretical point of view. On the basis of the largely accepted
magnetospheric accretion scenario, gas from the circumstellar disk
accretes onto the star surface following the magnetic field lines of the
stellar magnetosphere \citep{Koenigl1991ApJ}. Since the circumstellar disk
is truncated at $\sim 3-10$ stellar radii, accreting gas is accelerated up
to velocities of $\sim 200-600$ km s$^{-1}$ by the stellar gravitational
field and, when impacting on the high-density layers of the stellar
atmosphere, it forms a strong shock that heats up the accretion column to
temperatures of few MK \citep{1994A&A...287..131G, Lamzin1998ARep}. The
magnetospheric accretion paradigm is supported by various observational
evidence: optical emission lines with inverse P Cygni profiles
indicate infalling material (e.g. \citealt{Edwards1994AJ}), UV and
optical excess comes from a photosphere heated by the accretion shock
\citep{Calvet1998ApJ, Gullbring2000ApJ}, and transition emission lines
in the far-UV are produced by the shock-heated material at temperature
of about 10$^5$ K (e.g. \citealt{Ardila2002ApJ}, \citealt{Herczeg2005AJ}).

More recently, high resolution ($R \sim 600$) X-ray observations
of some CTTSs (TW Hya, BP Tau, V4046 Sgr, Hen 3-600, MP Mus
and RU Lupi) have revealed a soft X-ray excess produced
by plasma at temperature $T\sim 2-3 \times 10^{6}$ K and
electron number density $n_{\rm e}\sim 10^{11}-10^{13}$ cm$^{-3}$
\citep{Kastner2002ApJ,Schmitt2005A&A,Gunther2006A&A, Huenemoerder2007ApJ,
Argiroffi2007A&A, Robrade2007A&A}. This component could be produced
by the shock-heated plasma at the base of the accretion column,
the plasma density being much higher than that ($n_{\rm e}\leq 10^{10}$
cm$^{-3}$) measured for coronae of active stars \citep{Testa2004ApJ}. The
discovery of optical depths effects on the X-ray spectrum of MP Mus
further supports this hypothesis \citep{Argiroffi2009A&A}.

The idea that the soft X-ray excess detected in CTTSs is due to
shocks formed at the impact of accretion columns onto the stellar surface
has recently received a convincing theoretical support by time-dependent
models of radiative accretion shocks in CTTSs \citep{Koldoba2008MNRAS,
Sacco2008A&A, Orlando2009A&A}. In particular, these studies have predicted
post-shock plasma characterized by density and temperature values in
the range observed and global shock oscillations induced by radiative
cooling instabilities analogous to those predicted in other astrophysical
contexts (e.g. \citealt{1993ApJS...88..253S, 1996ApJS..102..161D,
1998ApJ...494..336S, 2003ApJ...591..238S, 2003ApJS..147..187S,
2005ApJ...626..373M}). Sacco et al. (2008; hereafter Paper I) developed
a one-dimensional hydrodynamic model with the aim to investigate the
dynamics and the stability of the shock-heated accreting material and
the role of the stellar chromosphere in determining the position and
the thickness of the shocked region. The model takes into account all
the important physical effects, i.e the gravity stratification, the
thermal conduction, the radiative losses from an optically thin plasma,
and a detailed description of the stellar chromosphere. For simulations
based on the parameters of MP~Mus, they synthesized the high resolution
X-ray spectrum, as it would be observed with the Reflection Grating
Spectrometers (RGS) on board the XMM-Newton satellite, and found an
excellent agreement with the observations \citep{Argiroffi2007A&A}.

Although the time-dependent models of radiative shocks support well the
origin of the soft X-ray excess in CTTSs, several observational
points remain still unclear: a) the mass accretion rates
derived from X-ray observations are generally lower than those
derived from optical and UV data by one or two orders of magnitude
(\citealt{Schmitt2005A&A, Gunther2007A&A, Argiroffi2009A&A}; Curran et
al. 2010, in prep.); b) some young accreting objects (T Tau, AB Aur,
HD 163296) do not exhibit spectral features indicating dense X-ray
emitting plasma \citep{Gudel2007A&A, Telleschi2007A&A, Gunther2009A&A};
c) a detailed time series analysis of the soft X-ray emission from
TW~Hydrae revealed no periodic variations \citep{Drake2009ApJ}. In
addition, the absorption from the stellar atmosphere could play an
important role in the observability of accretion shocks in X-rays,
especially for high density streams \citep{Drake2005ESASP}, but up to
now this effect has been not fully explored.

More recently, \cite{Brickhouse2010ApJ} performed a long exposure
X-ray observation of TW Hydrae, using the \textit{Chandra} High Energy
Trasmission Grating, that provides a rich set of diagnostics for electron
temperature, electron density and the hydrogen column density. Temperature
($T_{\rm e}\approx 2.5$ MK) and electron density ($n_{\rm e}\approx
3.0\times 10^{12}$ cm$^{-3}$) derived from the He-like \ion{Ne}{ix}
line ratio diagnostics are in agreement with a standard accretion
shock model describing a single stream with uniform density and
velocity. However, this model cannot explain the lower values of
plasma density ($n_{\rm e}\approx 6.0 \times 10^{11}$ cm$^{-3}$) and
temperature ($T_{\rm e}\approx 1.5$ MK) derived from the \ion{O}{vii}
line ratio, because single stream models predict that the density
of the post-shock plasma increases as the temperature decreases. In
order to explain both the \ion{O}{vii} and \ion{Ne}{ix} diagnostics,
\cite{Brickhouse2010ApJ} suggested that the X-ray emission originates
from three plasma components: a hot (T$_{\rm e}\approx 10$ MK) corona, a
high density ($n_{\rm e}\approx 6.0\times 10^{12}$ cm$^{-3}$, $T_{\rm
e}\approx 3.0$ MK) post-shock region close to the shock surface, and a
cold less dense ($n_{\rm e}\approx 2\times 10^{11}$ cm$^{-3}$, $T_{\rm
e}\approx 2.0$ MK) post-shock cooling region, with 300 times more volume
and 30 times more mass than that of the post shock region itself.

In this paper we analyze the observability of the X-ray emission from the
post-shock plasma as a function of the properties of the accretion stream
(density, velocity, and metal abundance). The main targets of our work
are: a) determining the main signatures of the accretion shock in the
X-ray band; b) estimating how the X-ray luminosity from the shock-heated
plasma depends on the properties of the accretion stream; c) investigating
the influence of the absorption from the stellar atmosphere on the
observed X-ray emission from the post-shock zone; d) understanding if
the discrepancy between accretion rates measured from optical and X-ray
data as well as the different densities measured from the \ion{O}{vii}
and \ion{Ne}{ix} He-like ions can be explained in the framework
of 1-D hydrodynamic models, assuming plasma-$\beta \ll 1$. To answer
these questions, we perform a set of 1-D hydrodynamical simulations
using the model introduced in Paper I. We explore the parameter space
of the accretion stream, namely its mass density, velocity, and metal
abundance. From the results of the simulations, we synthesize the
X-ray emission arising from the post-shock zone in the $[0.5-8.0]$
keV band and in the resonance lines of the He-like ions \ion{O}{vii}
and \ion{Ne}{ix}, taking into account the effect of the absorption from
the surrounding stellar chromosphere.

The paper is organized as follow: Sect. \ref{sec2} describes the
numerical setup and the space of the physical parameters explored
by the simulations; Sect. \ref{sec3} reports the results focusing
on the properties of the post-shock zone and on the X-ray emission;
Sect. \ref{sec4} discusses our results compared with the properties
of CTTSs observed by high resolution X-ray spectrographs; in Sect.
\ref{sec5}, we draw our conclusions.

\section{The Model}
\label{sec2}

\subsection{The numerical setup}

\begin{figure}[bt]
\includegraphics[width=8.5cm]{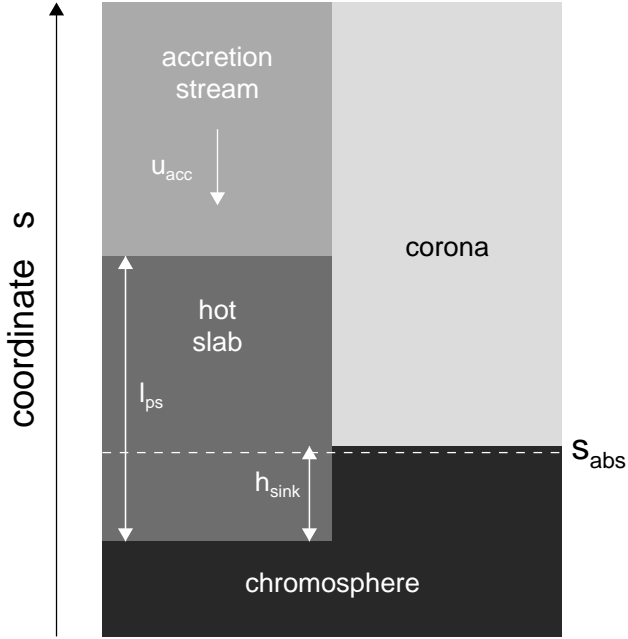}
\caption{ Schematic description of the system geometry. $l_{ps}$ and $h_{sink}$
are the expected thickness of the post-shock zone and the distance of the base of the slab 
from the transition region between the chromosphere and the stellar corona, respectively
(see Sect. \ref{par:spacepar}). $s_{abs}$ is the threshold below which, we assume the 
X-ray emission is fully absorbed (see Sect. \ref{sect:absorption}). 
}
\label{fig:cartoon}
\end{figure}

We adopt the model introduced in Paper I. We assume that accretion
occurs along the magnetic flux tube linking the circumstellar disk to the
star and that plasma moves and transports energy only along the magnetic
field lines. This assumption is valid for values of the plasma parameter
$\beta \ll 1$ (where $\beta =$ gas pressure / magnetic pressure) in the
accretion column. We assume the accretion streams to be perpendicular to
the stellar surface and focus our analysis on the portion of the stream
close to the star with a constant mass accretion rate. This allows us to
assume the plane-parallel geometry (the maximum expected ratio between 
the hot slab thickness and the stellar radius is $<0.2$). 
 A schematic description of the system geometry is shown in 
Fig. \ref{fig:cartoon}. The model is one-dimensional and describes the 
impact of the stream onto the chromosphere along the coordinate s 
(see Fig. \ref{fig:cartoon}).  The plasma dynamics is described by solving 
numerically the time-dependent fluid equations
of mass, momentum, and energy conservation for a compressible conducting
and optically thin plasma fluid above the stellar surface:

\begin{equation}
\frac{\partial \rho}{\partial t} +  \frac{\partial \rho
u}{\partial s} = 0~,
\label{eq:massa-1}
\end{equation}

\begin{equation}
\frac{\partial \rho u}{\partial t} +\frac{\partial
(P+\rho u^2)}{\partial s} = \rho g~,
\label{eq:momento-1}
\end{equation}

\begin{equation}
\frac{\partial \rho E}{\partial t} +\frac{\partial
(\rho E+P) u} {\partial s} = \rho u g + E_{H} -  \frac{\partial
q}{\partial s} - n_{\rm e} n_{\rm H} \Lambda(T)~,
\label{eq:en+r+c-1}
\end{equation}

\[
\epsilon =\frac{P}{\rho (\gamma-1)}~,\hspace{0.8cm}
P=\left[1+\alpha(\rho,T)\right]\frac{\rho  k_{\rm B} T}{\mu m_H},
\]

\noindent 
where $t$ is the time; $s$ is the coordinate along the magnetic field
lines  (see Fig. \ref{fig:cartoon}); $u$ is the plasma velocity; $\rho = \mu m_{\rm H} n_{\rm H}$ is
the mass density; $\mu$ is the mean atomic mass, that ranges between
1.277 and 1.381 times hydrogen mass as function of the adopted metal
abundance; $m_{\rm H}$ is the mass of the hydrogen atom; $n_{\rm e}$
and $n_{\rm H}$ are the electron number density and the hydrogen
number density, respectively; $P$ is the thermal pressure; $g(s)$ is
the gravity of a star with a mass $M=0.8~{M_{\sun}}$ and a radius
$R=1.3~{R_{\sun}}$; $T$ is the plasma temperature; $E=\epsilon +
u^2/2$ is the total gas energy per unit mass; $\epsilon$ is the internal
energy per unit mass; $\gamma=5/3$ is the ratio between specific heats;
$k_{\rm B}$ is the Boltzmann constant; $q$ is the heat flux; $\Lambda(T)$
is the radiative losses function per unit emission measure; $\alpha(\rho,
T)$ is the fractional ionization ($n_{\rm e}/n_{\rm H}$), that has
been derived from a modified Saha equation for the solar chromosphere
conditions \citep{Brown1973SoPh}; $E_{\rm H}$ is a parametrized
chromospheric heating function ($E_{\rm H} =0$ for $T> 8 \times
10^3$ K) defined, as in \cite{Peres1982ApJ}, to keep the unperturbed
chromosphere in stable equilibrium.

The thermal conduction includes a smooth transition between the classical
and the saturated conduction regime, as in \cite{Dalton1993ApJ}

\begin{equation} q =
\left(\frac{1}{q_{\rm spi}}+\frac{1}{q_{\rm sat}}\right)^{-1}
\end{equation}

\noindent
where $q_{\rm spi}=-\kappa(T)\partial T /\partial s$ is the classical
conductive flux \citep{Spitzer1962} (with $\kappa(T)= 9.2 \times
10^{-7}T^{5/2}$ erg s$^{-1}$ K$^{-1}$ cm$^{-1}$), $q_{\rm sat} = -{\rm
sign}\,( \partial T /\partial s)\phi\rho c_{\rm s}^3$ is the saturated
flux \citep{Cowie1977ApJ}, $\phi\leq 1$ (\citealt{Borkowski1989ApJ}
and references therein) and $c_{\rm s}$ is the isothermal sound speed.

The radiative losses per unit of emission measure $\Lambda(T)$ have been
calculated with the PINTofALE spectral code \citep{Kashyap2000BASi}
and the APED V1.3 atomic line database \citep{Smith2001ApJ}. Since
the radiative losses in the temperature range $10^5 < T < 10^7$ K are
dominated by the emission lines from heavy ions, the radiative cooling
depends on the metal abundance, $\zeta$, which is one of the parameters
explored in this work.

The equations are solved numerically using the \FLASH\ code
\citep{Fryxell2000ApJS} with its implementation of the PPM algorithm
\citep{Colella1984JCoPh}. \FLASH\ is a multi-physics code for solving
astrophysical problems that handles an adaptive mesh by the \PARAMESH\
algorithm \citep{MacNeice2000CoPhC}. The code has been extended with an
additional module for the evolution of the fractional hydrogen
ionization ($n_{\rm e}/n_{\rm H}$).

The extension of the computational domain depends on the specific
parameters of the simulations (see Sect. \ref{par:spacepar}). The largest
domain considered here extends over a range $D = 2.4\times 10^{10}$ cm
above the stellar surface, for the simulation that analyzes an accretion
stream density $n_{\rm e}=10^{10}$ cm$^{-3}$, velocity $u_{\rm acc}=600$
km~s$^{-1}$, and solar abundances (see Sect. \ref{par:spacepar} for a
detailed description of the space of the physical parameters explored).

The spatial resolution adopted for each simulation depends on
the thickness of the post-shock zone that spans a range of 6 orders of
magnitude (See Sect. \ref{sec:post-shock}). At the highest resolution
(stream with density $n_{\rm e}=10^{13}$ cm$^{-3}$, velocity $u_{\rm
acc}=200$ km s$^{-1}$ and metal abundance $\zeta = 5$, in solar units)
we allow for a maximum of 16 levels of refinement in the \PARAMESH\
algorithm, with resolution increasing twice at each refinement level,
and with the refinement criterion following the changes in density
and temperature. This grid configuration yields an effective maximum
resolution of $\approx 10^3$ cm at the finest level. For the simulation
requiring the highest spatial resolution, we analyzed the effect of
resolution on the model solution by considering additional simulations
which use an identical setup but with a larger number of refinement
levels. We checked that the resolution adopted in this paper is the best
compromise between accuracy and computational cost and that the system
evolution is well described in its details.

\begin{figure}[h]
\includegraphics[width=8 cm]{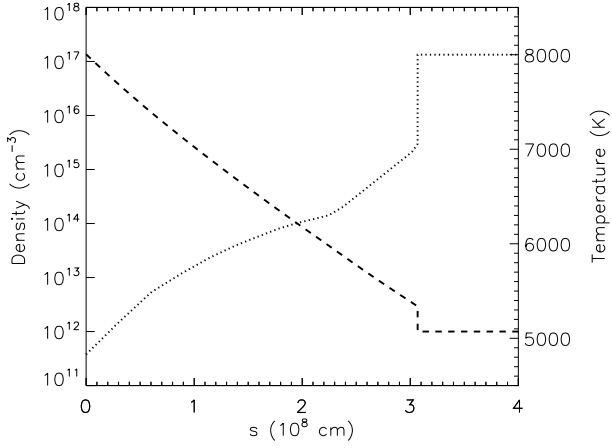}
\caption{Density (dashed line) and temperature (dotted line) profiles for
the initial unperturbed stellar chromosphere ($s < 3.05\times 10^8$ cm)
and an accretion stream ($s > 3.05\times 10^8$ cm) with hydrogen number
density $n_{\rm acc}=10^{12}~cm^{-3}$ and solar metal abundance.}
\label{fig:initial_cond}
\end{figure}

As initial condition we consider a stream with uniform density,
temperature and velocity impacting onto a hydrostatic chromosphere (see
Fig.~\ref{fig:initial_cond}). The temperature of the stream is the same in
all the simulations performed and does not influence the system evolution,
because the thermal energy of the stream is negligible with respect
to its kinetic energy. The chromosphere of the young star is described
adopting the solar models in \cite{Vernazza1973ApJ}, rescaled to have,
at its base, a pressure ($\approx 9 \times 10^4$ dyn cm$^{-2}$) larger
than the ram pressure of the streams considered here and an extension
of $3 \times 10^8$ cm. As boundary conditions, the values of density,
temperature, and velocity are fixed to the initial ones both at the top
(i.e. a constant accretion rate) and at the base of the computational
domain. We checked that the adopted chromosphere and boundary conditions
prevent boundary effects on the system evolution.

\subsection{Space of the physical parameters}
\label{par:spacepar}

We explore the space defined by three parameters: the velocity and
density of the pre-shock accretion flow, and the metal abundance of
both the stellar and the accreting matter, that are assumed to be
equal.  Note however that a discrepancy between chromospheric and accretion
metal abundances would not affect the results of our simulations, because, 
as discussed in \cite{Sacco2008A&A} and in the Sect. \ref{sec:post-shock} of this paper, 
the chromosphere acts as reservoir of matter stopping the accretion flow, 
but it does not influence the properties of the post-shock zone. 
Instead, different chromospheric abundances can
slightly affect the absorption from the chromosphere as discussed in Sect. \ref{sect:absorption}.
We consider three pre-shock velocities ($u_{\rm
acc}=200,~400,~600$ km~s$^{-1}$), 4 pre-shock densities ($n_{\rm
acc}=10^{10},~10^{11},~10^{12},~10^{13}$ cm$^{-3}$) and 3 metal
abundances ($\zeta = 0.2, 1.0, 5.0$ in solar units retrieved from
\citealt{Anders1989GeCOA}). Both densities and metal abundances span 
larger ranges than those obtained from high spectral resolution 
X-ray observations. We focused our analysis on a large space of parameters
for two main reasons. Probably the most important is that current high 
spectral resolution  X-ray observations of CTTSs
are limited to a very small sample of very close and bright CTTSs. 
These stars are older and accrete to a lower rate than typical CTTSs and
it is most likely that physical and chemical properties of the 
corresponding accretion stream may differ too. Indeed, \cite{Bary2008ApJ} and 
\cite{Martin1996ApJ}
estimated pre-shock densities $n_{\rm acc}=10^{12}-10^{13}~cm^{-3}$, while 
pre-shock densities 
derived from X-ray data are $n_{\rm acc}=10^{11}-10^{12}~cm^{-3}$. 
Moreover, densities and metal abundances are measured from the spatially integrated 
emission including both the coronal and the accretion component. 
Typical coronal densities ($n_{e}<10^{10}~cm^{-3}$) are much lower  
than densities expected in the accretion stream and metal abundances
are lower than photospheric abundances observed in young stars
\citep{Telleschi2007A&A}.
Therefore, both accretion stream densities and abundances derived from 
X-ray observations could be underestimated.

A heuristic model that assumes the strong shock approximation
\citep{Zeldovich1967book}, stationary conditions and radiative cooling
has been previously used to describe accretion shock physics in CTTSs
\citep{Lamzin1998ARep, Calvet1998ApJ} and to derive mass accretion rates
from the X-ray emission \citep{Argiroffi2007A&A}. This model provides
post-shock region characteristics:

\begin{equation}
u_{\rm ps}=\frac{u_{\rm acc}}{4}~, \hspace{0.8 cm} n_{\rm ps}=4 n_{\rm acc}~,
\label{eqn:vel_dens}
\end{equation}

\begin{equation}
\tau_{\rm cool}= \frac{1}{\gamma-1}\frac{P}{n_{\rm e} n_{\rm H} \Lambda
(T)}\sim 2.5 \times 10^3\, \frac{1}{\zeta}\,
\frac{T_{\rm ps}^{3/2}}{n_{\rm ps}}~,
\label{eqn:taucool}
\end{equation}

\begin{equation}
\tau_{\rm cross}=\frac{l_{\rm ps}}{u_{\rm ps}} \equiv \tau_{\rm cool}~,
\end{equation}

\begin{equation}
T_{\rm ps}=\frac{3}{32}\frac{\mu m_H}{k_{\rm B}}u_{\rm acc}^2 \approx 1.4\times
10^{-9} u_{\rm acc}^2~,
\label{eqn:posttemp}
\end{equation}

\begin{equation}
l_{\rm ps} \equiv \tau_{\rm cool} u_{\rm ps} = 1.7\times 10^{-11} \,
\frac{1}{\zeta}\, \frac{u_{\rm acc}^{4}}{n_{\rm acc}}~,
\label{eqn:lslab}
\end{equation}

\noindent
where $u_{\rm ps}$ and $n_{\rm ps}$ are the post-shock velocity and
density, $u_{\rm acc}$ and $n_{\rm acc}$ are the stream pre-shock velocity
and density, $\tau_{\rm cross}$ is the crossing time of the accreting
material through the post-shock zone (which is expected to
be equal to the radiative cooling time $\tau_{\rm cool}$), $T_{\rm ps}$
is the post-shock temperature, and $l_{\rm ps}$ is the thickness of the
post-shock zone, and where the radiative cooling function has been
approximated as $\Lambda(T)\approx 1.6 \times 10^{-19} \zeta T^{-1/2}$
erg~s$^{-1}$~cm$^{3}$ (e.g. \citealt{Orlando2005A&A}).

\setcounter{table}{0}
\begin{table}[!t]
\caption{Range of relevant physical parameters explored.}
\label{tab:Par_space}
\begin{center}
\begin{tabular}{ccc}
\hline
\hline
Parameter          & Range                     & Units \\ 
\hline
$n_{\rm acc}$      &  $10^{10}-10^{13}$        &   cm$^{-3}$   \\
$u_{\rm acc}$      &  $200-600$                &   km s$^{-1}$   \\
$\zeta$            &  $0.2-5.0$                &        \\
\hline
$T_{\rm ps}$       &  $5.8 \times 10^5-5.6 \times 10^6$  &   K   \\
$\tau_{\rm cool}$  &  $8.3 \times 10^{-3}-1000$  &   s   \\
$l_{\rm ps}$       &  $4.2 \times 10^4-1.5\times10^{10}$   &   cm   \\
$P_{\rm ram}$      &  $35-7.8\times10^4$       &   dyn cm$^{-2}$   \\
$h_{\rm sink}$     &  $1.3\times 10^8-3.5\times 10^8$     &   cm   \\
$\beta_{\rm ps}$   &  $8.6\times10^{-4}-2.0$   &        \\
\hline 
\end{tabular} 
\end{center}
\smallskip
$n_{\rm acc}$ is the stream hydrogen number density; $u_{\rm acc}$ is the
stream velocity; $\zeta$ is the heavy element abundance; $T_{\rm ps}$ is
the post-shock temperature; $\tau_{\rm cool}$ is the radiative cooling
time; $l_{\rm ps}$ is the expected thickness of the post-shock zone;
$P_{\rm ram} = \rho_{\rm acc} u_{\rm acc}^2$ is the ram pressure of the
accretion flow; $h_{\rm sink}$ is the sinking of the post-shock zone in
the chromosphere; $\beta_{\rm ps} =P_{\rm ram}/(B^2/8\pi)$ is the plasma
parameter in the post-shock zone, assuming a magnetic field $B=1$ kG.
\end{table}

Table \ref{tab:Par_space} reports the ranges of values of all the
relevant physical parameters of the simulations explored here. The
first three rows report the three independent parameters of our analysis
(i.e. density, velocity, and metal abundance of the stream), whereas the
following rows report the parameters derived from the independent ones,
namely the post-shock temperature $T_{\rm ps}$ (Eq. \ref{eqn:posttemp}),
the radiative cooling time $\tau_{\rm cross}$ (Eq. \ref{eqn:taucool}), 
the slab thickness $l_{\rm ps}$ (Eq. \ref{eqn:lslab}  and Fig. \ref{fig:cartoon}), 
the ram pressure of the accretion stream $P_{\rm ram}=\rho_{\rm acc} u_{\rm acc}^2$,
the sinking of the slab in the chromosphere $h_{\rm sink}$ (i.e. the
distance of the base of the slab from the transition region between
the chromosphere and the stellar corona,  see Fig. \ref{fig:cartoon}), 
and the plasma parameter of the post-shock zone $\beta=P_{\rm ram}/(B^2/8\pi)$, 
where the magnetic field strength is assumed to be $B\sim 1$ kG (which is the
order of magnitude of the field strengths derived from observations
of CTTSs; \citealt{Johns-Krull2007ApJ}). Note that the plasma $\beta$
is $\ll 1$ for most of the accretion streams considered in this work
(thus justifying the low-$\beta$ assumption of our model) except for 5
cases, where $\beta \sim 1-2$ (for the magnetic field assumed). On the
other hand, \cite{Orlando2009A&A} showed that, for $\beta \sim 1-5$,
the evolution of radiative shocks in 2D MHD models (thus including an
explicit description of the ambient magnetic field) is analogous to
that described by 1-D hydrodynamic models, although the amplitude of
oscillations is smaller and the frequency higher than those predicted by
1-D models. For the 30 different cases analyzed here, the duration of the
simulations ranges from 100 to 19000 s. The duration of each simulation
was set much longer than the initial transient effect, at the impact
of the stream onto the chromosphere, and to include a large number (at
least three) of oscillation periods of the post-shock zone (see Paper I).

\subsection{Synthesis of the X-ray emission \label{sect:X-ray_syn}}

For each simulation, we synthesize the X-ray luminosity ($L_{\rm X}$)
in the $[0.5-8.0]$ keV energy band and in the resonance lines of two
He-like ions, namely the \ion{O}{vii} at 21.60 {\AA} ($L_{\rm OVII}$)
and the \ion{Ne}{ix} at 13.45 {\AA} ($L_{\rm NeIX}$), which can
be measured with the high resolution spectrographs on board the
XMM-{\it Newton} and {\it Chandra} satellites. In particular:

\begin{enumerate}

\item From the spatial distributions of density and temperature (output of
the numerical simulations), we compute the emission measure of the plasma
in the $j$th domain cell as ${\rm em}_{\rm j} = n_{\rm Hj}^2 A_{\rm str}
\Delta s_{\rm j}$, where $n_{\rm Hj}$ is the particle number density
in the cell, $\Delta s_j$ is the length of the $j$th cell along the
stream, and $A_{\rm str}$ is the cross section of the stream. In this
paper we assume $A_{\rm str}=5\times 10^{20}$ cm$^2$, (i.e. $\sim 0.5$\%
of the stellar surface); for the values of stream velocity and density
explored here, this cross section produces mass accretion rates spanning
the range of values derived from X-ray observations of CTTSs (see Sect.
\ref{sec:high_energy}). We then derive the distribution of emission
measure versus temperature, EM($T$), by binning the emission measure
values into slots of temperature; the range of temperature [$4 < \log T
(\mbox{K}) < 8$] is divided into 81 bins, all equal on logarithmic scale
($\Delta \log(T)=0.05$).

\item From the EM($T$) distributions, we synthesize the X-ray emission
in the $[0.5-8.0]$ keV band and in the resonance lines of \ion{O}{vii}
and \ion{Ne}{ix}, considering the appropriate metal abundance $\zeta$,
using the PINTofALE spectral code with the APED V1.3 atomic line database.

\item Finally, we derive time-average luminosities $L_{\rm X}$, $L_{\rm
OVII}$, $L_{\rm NeIX}$ over a selected time interval including, at least,
three oscillation cycles and outside initial transients.

\end{enumerate}

\subsection{Absorption from the stellar atmosphere}
\label{sect:absorption}

As pointed out by \cite{Drake2005ESASP}, a large fraction of the X-ray
emission due to the accretion shock may be absorbed by the gas present
in the surrounding stellar atmosphere or by the accretion column itself.
The absorption from the stellar atmosphere mainly depends on the location
of the emitting plasma. For a given shocked slab of material rooted
in the chromosphere, photons emitted from plasma located at the base
of the slab (deep into the chromosphere) travel through higher density
gas layers than photons from plasma located in the shallower portion
of the slab. The deepness of the emitting plasma in the chromosphere
depends on the thickness of the slab $l_{\rm ps}$ and on the sinking of
the slab in the chromosphere $h_{\rm sink}$ (Sect. \ref{par:spacepar}) 
to the position  at which the ram pressure of the post-shock plasma equals
the thermal pressure of the chromosphere. As discussed in Sect. 
\ref{par:spacepar}, our simulations explore a wide range of values for 
$l_{\rm ps}$ and $h_{\rm sink}$ (see Table~\ref{tab:Par_space}).

In order to estimate the effect of the absorption on the observed X-ray
emission, we calculate the X-ray luminosities by assuming that only
the emission from plasma located in the shallower layers of the shocked
slab can be observed (see also \citealt{Drake2005ESASP} for a previous
application of this method to estimate absorption effects). Specifically,
we assume that all the emission due to plasma located at $s<s_{\rm abs}$
is fully absorbed, while the emission produced from plasma located
at $s>s_{\rm abs}$ is fully transmitted. The threshold $s_{\rm abs}$
 is defined as the height, in the unperturbed stellar chromosphere, at
which the overlying atmosphere absorbs 50\% of the energy of a trial
X-ray spectrum (see also Fig. \ref{fig:cartoon}). We considered 
as trial X-ray spectrum that produced by a plasma at $T=1$\,MK.
 The plasma temperature does not influence the threshold $s_{\rm abs}$ used for 
calculating the \ion{Ne}{ix} and the \ion{O}{viii} resonance line luminosities, 
while it slightly
influences the thresholds used for the $[0.5-8.0]$\,keV band.
However, we found that the results of only a few of the parameter 
configurations considered here,
specifically those leading to $l_{ps}\sim h_{sink}$, slightly depend on 
the particular choice of the trial X-ray spectrum. 
For these configurations a more detailed description of the absorption effect,
including the dependence on the wavelength, a more smooth transition between
the full absorption and the full transmission and a dependence of the angle of view
should be performed to give a more precise answer to the observability issue.

\setcounter{table}{1}
\begin {table}[!t]
\caption{\label{tab:threshold} Values of the threshold $s_{\rm abs}$
as a function of the energy band and of the metal abundance.}
\centering
\begin{tabular}{cccc}
\hline
\hline
$\zeta$ & $[0.5-8.0]$ keV  & \ion{O}{vii} & \ion{Ne}{ix} \\ 
        &   ($10^8$ cm)      &  ($10^8$ cm)        &   ($10^8$ cm)      \\
\hline
0.2   & 3.56  & 3.57   & 3.17    \\
1.0   & 3.72  & 3.74   & 3.43   \\
5.0   & 3.81  & 3.83   & 3.58   \\
\hline 
\end{tabular} 
\end{table}

\begin{figure*}[!ht]
\sidecaption
\includegraphics[width=12 cm]{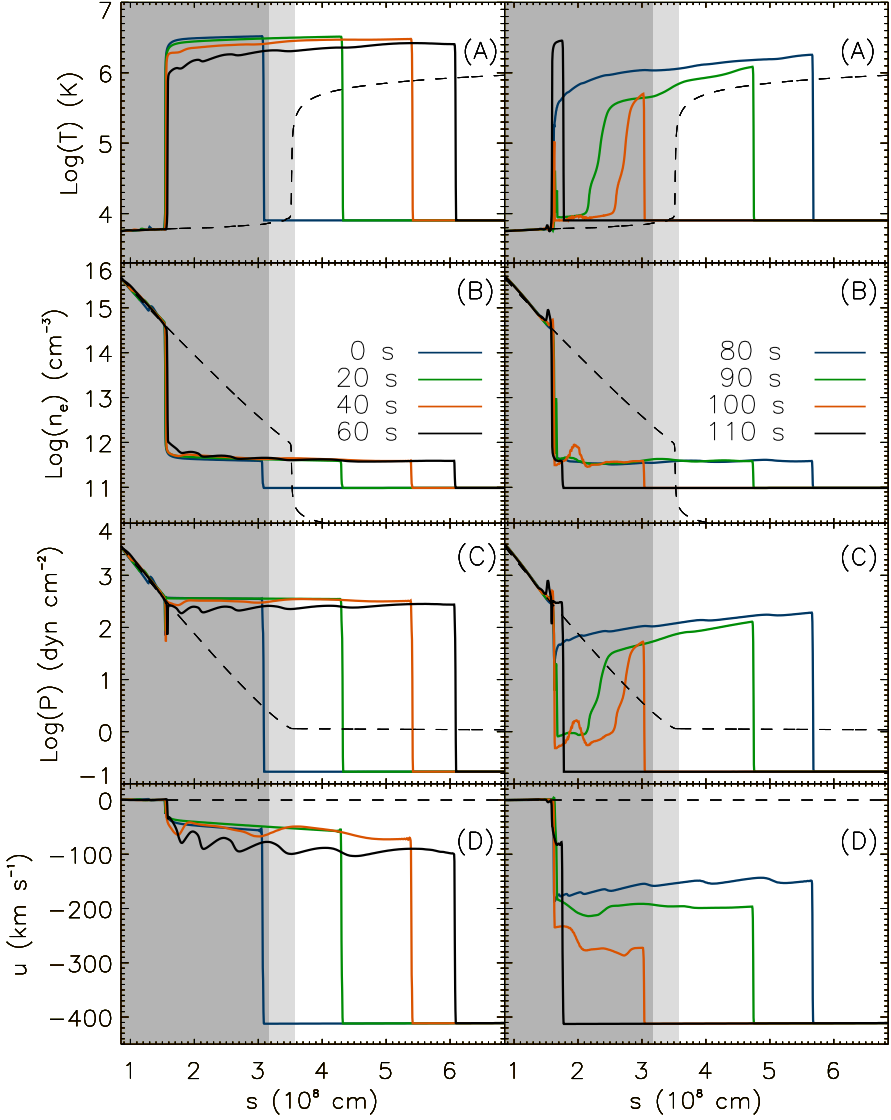}
\caption{Time evolution of plasma temperature (A), density (B), pressure
(C) and velocity (D) during a complete oscillation of the post-shock zone
in the case of an accretion stream with density $n_{\rm acc}=10^{11}$
cm$^{-3}$, velocity $u_{\rm acc}=400$ km~s$^{-1}$ and metal abundance
$\zeta=1.0$. The left (right) panels show the time evolution during
the heating (cooling) phase. The dashed lines describe the profiles of
the unperturbed initial stellar atmosphere (chromosphere and corona). The
shaded areas mark the regions below the thresholds $s_{\rm abs}$
used to estimate the effects of absorption on the luminosity in the
\ion{O}{vii} (pale grey) and \ion{Ne}{ix} resonance lines (see Sect.
\ref{sect:absorption}).}
\label{fig:oscillations}
\end{figure*}

The values of $s_{\rm abs}$ for the three X-ray bands and for the three metal
abundances are reported in Table \ref{tab:threshold}.
The value of $s_{\rm abs}$ obviously depends on the chromospheric 
metal abundances, that we assume to be equal to the abundances of the
accreting material. However, chromospheric abundances may be different
from accretion stream abundances. For instance,
if the chromospheric abundance is higher or lower than the accretion
stream abundance, then we underestimate or overestimate the absorption effect,
respectively. A better knowledge of metal abundances of the
different components of the star-disk system is required to better
address this issue.


We found that $s_{\rm abs}$ does not change significantly if we consider
the absorbed X-ray emission in the $[0.5-8.0]$\,keV band or in the
\ion{O}{vii} resonance line, because the absorption in the softer part
of the $[0.5-8.0]$\,keV band (which is the same of the \ion{O}{vii}
line) is dominant. On the other hand, we found a significantly lower
value of $s_{\rm abs}$ in the \ion{Ne}{ix} resonance line, this
line forming at higher energy (where the absorption is lower) than
\ion{O}{vii}. Therefore, we consider one set of thresholds $s_{\rm abs}$
for the synthesis of the luminosities in the $[0.5-8.0]$\,keV band and
in the \ion{O}{vii} resonance line, and another set of $s_{\rm abs}$
for the luminosities in the \ion{Ne}{ix} resonance line (see Table
\ref{tab:threshold}).

\section{Results}
\label{sec3}

\subsection{Physical properties of the post-shock zone
\label{sec:post-shock}}

\setcounter{table}{2}
\begin {table*}[!t]
\caption{Physical properties of the post-shock zone and X-ray luminosities
}
\centering
\begin{tabular}{cccccccccccc}
\hline
\hline
  $n_{\rm acc}$ & $u_{\rm acc}$ & $\zeta$ &
 $P_{\rm osc}$ & $l_{\rm max}$ & $T_{\rm ps}$ & $\log(L_{\rm X})^{\,a}$
& $\log(L_{\rm X})^{\,b}$
& $\log(L_{\rm OVII})^{\,c}$ & $\log(L_{\rm OVII})^{\,d}$ & $\log(L_{\rm
NeIX})^{\,e}$ & $\log(L_{\rm NeIX})^{\,f}$ \\
  (cm$^{-3}) $ & (km s$^{-1}$)&  &  (s) & (cm) & (MK) & (erg s$^{-1}$)
& (erg s$^{-1}$) & (erg s$^{-1}$) & (erg s$^{-1}$) & (erg s$^{-1}$) &
(erg s$^{-1}$) \\
\hline
10$^{10}$ & 400 & 1.0 & 1.3$\times 10^{ 3}$ & 5.0$\times 10^{ 9}$ & 2.3 &   28.41 &   28.40 &   27.60 & 27.60 & 26.79 & 26.79 \\
10$^{10}$ & 400 & 5.0 & 5.0$\times 10^{ 2}$ & 6.0$\times 10^{ 8}$ & 2.0 &   28.20 &   28.11 &   27.58 & 27.46 & 26.45 & 26.43 \\
10$^{10}$ & 600 & 1.0 & 3.8$\times 10^{ 3}$ & 1.5$\times 10^{10}$ & 4.1 &   29.32 &   29.32 &   28.04 & 28.00 & 27.66 & 27.66 \\
10$^{10}$ & 600 & 5.0 & 1.2$\times 10^{ 3}$ & 5.0$\times 10^{ 9}$ & 4.6 &   29.51 &   29.48 &   28.11 & 28.00 & 27.81 & 27.79 \\
10$^{11}$ & 200 & 0.2 & 6.1$\times 10^{ 1}$ & 8.0$\times 10^{ 7}$ & 0.5 &   25.52 &   25.30 &   24.90 & 24.68 & 21.67 & 21.67 \\
10$^{11}$ & 200 & 1.0 & 1.5$\times 10^{ 1}$ & 2.0$\times 10^{ 7}$ & 0.5 &   25.75 &      -  &   25.18 &    -  & 22.08 & 21.36 \\
10$^{11}$ & 200 & 5.0 & 3.2$\times 10^{ 0}$ & 4.2$\times 10^{ 6}$ & 0.5 &   25.67 &      -  &   25.11 &    -  & 21.97 &    -  \\
10$^{11}$ & 400 & 0.2 & 5.0$\times 10^{ 2}$ & 1.5$\times 10^{ 9}$ & 2.1 &   29.30 &   29.26 &   28.53 & 28.49 & 27.56 & 27.54 \\
10$^{11}$ & 400 & 1.0 & 1.6$\times 10^{ 2}$ & 4.0$\times 10^{ 8}$ & 2.1 &   29.18 &   28.89 &   28.49 & 28.18 & 27.46 & 27.32 \\
10$^{11}$ & 400 & 5.0 & 4.2$\times 10^{ 1}$ & 5.5$\times 10^{ 7}$ & 2.1 &   29.00 &      -  &   28.36 &    -  & 27.23 &    -  \\
10$^{11}$ & 600 & 0.2 & 1.3$\times 10^{ 3}$ & 7.0$\times 10^{ 9}$ & 4.4 &   30.36 &   30.34 &   28.89 & 28.85 & 28.54 & 28.53 \\
10$^{11}$ & 600 & 1.0 & 5.0$\times 10^{ 2}$ & 2.5$\times 10^{ 9}$ & 4.6 &   30.40 &   30.34 &   28.96 & 28.88 & 28.65 & 28.61 \\
10$^{11}$ & 600 & 5.0 & 3.8$\times 10^{ 2}$ & 3.7$\times 10^{ 8}$ & 4.9 &   30.18 &   29.72 &   28.59 & 27.72 & 28.51 & 28.04 \\
10$^{12}$ & 200 & 0.2 & 5.7$\times 10^{ 0}$ & 6.2$\times 10^{ 6}$ & 0.5 &   26.30 &      -  &   25.69 &    -  & 22.38 &    -  \\
10$^{12}$ & 200 & 1.0 & 1.4$\times 10^{ 0}$ & 1.7$\times 10^{ 6}$ & 0.5 &   26.54 &      -  &   25.96 &    -  & 22.76 &    -  \\
10$^{12}$ & 200 & 5.0 & 2.8$\times 10^{-1}$ & 3.7$\times 10^{ 5}$ & 0.5 &   26.51 &      -  &   25.92 &    -  & 22.66 &    -  \\
10$^{12}$ & 400 & 0.2 & 4.7$\times 10^{ 1}$ & 5.0$\times 10^{ 7}$ & 1.9 &   30.08 &      -  &   29.46 &    -  & 28.18 & 26.28 \\
10$^{12}$ & 400 & 1.0 & 1.2$\times 10^{ 1}$ & 1.6$\times 10^{ 7}$ & 1.9 &   30.18 &      -  &   29.54 &    -  & 28.43 &    -  \\
10$^{12}$ & 400 & 5.0 & 2.9$\times 10^{ 0}$ & 5.0$\times 10^{ 6}$ & 2.0 &   29.68 &      -  &   29.08 &    -  & 27.88 &    -  \\
10$^{12}$ & 600 & 0.2 & 3.3$\times 10^{ 2}$ & 5.5$\times 10^{ 8}$ & 4.4 &   31.11 &   30.72 &   29.62 & 28.91 & 29.36 & 29.04 \\
10$^{12}$ & 600 & 1.0 & 1.5$\times 10^{ 2}$ & 1.5$\times 10^{ 8}$ & 4.6 &   31.11 &      -  &   29.56 &    -  & 29.48 &    -  \\
10$^{12}$ & 600 & 5.0 & 2.9$\times 10^{ 1}$ & 2.5$\times 10^{ 7}$ & 4.8 &   31.08 &      -  &   29.49 &    -  & 29.43 &    -  \\
10$^{13}$ & 200 & 0.2 & 5.9$\times 10^{-1}$ & 3.5$\times 10^{ 5}$ & 0.5 &   27.43 &      -  &   26.81 &    -  & 23.60 &    -  \\
10$^{13}$ & 200 & 1.0 & 1.4$\times 10^{-1}$ & 1.5$\times 10^{ 5}$ & 0.5 &   27.15 &      -  &   26.53 &    -  & 23.11 &    -  \\
10$^{13}$ & 200 & 5.0 & 2.8$\times 10^{-2}$ & 3.2$\times 10^{ 4}$ & 0.5 &   27.49 &      -  &   26.91 &    -  & 23.63 &    -  \\
10$^{13}$ & 400 & 0.2 & 4.1$\times 10^{ 0}$ & 5.0$\times 10^{ 6}$ & 1.8 &   31.11 &      -  &   30.43 &    -  & 29.26 &    -  \\
10$^{13}$ & 400 & 1.0 & 2.2$\times 10^{ 0}$ & 1.5$\times 10^{ 6}$ & 2.1 &   30.91 &      -  &   30.32 &    -  & 29.04 &    -  \\
10$^{13}$ & 400 & 5.0 & 3.2$\times 10^{ 0}$ & 3.4$\times 10^{ 5}$ & 1.9 &   30.63 &      -  &   30.04 &    -  & 28.80 &    -  \\
10$^{13}$ & 600 & 0.2 & 8.3$\times 10^{ 1}$ & 3.8$\times 10^{ 7}$ & 4.4 &   32.08 &      -  &   30.45 &    -  & 30.36 &    -  \\
10$^{13}$ & 600 & 1.0 & 7.3$\times 10^{ 1}$ & 1.3$\times 10^{ 7}$ & 4.4 &   32.23 &      -  &   30.76 &    -  & 30.64 &    -  \\
\hline
\hline 
\multicolumn{12}{l}{$a$: Luminositity in the 0.5-8.0 keV band not considering the absorption.}\\
\multicolumn{12}{l}{$b$: Luminositity in the 0.5-8.0 keV band considering the absorption from the
stellar atmosphere as discussed in the text.}\\
\multicolumn{12}{l}{$c$: Luminositity in the \ion{O}{vii} resonance line at 21.6 \AA~not considering the absorption.}\\
\multicolumn{12}{l}{$d$: Luminositity in the \ion{O}{vii} resonance line at 21.6 \AA~considering the absorption from the
stellar atmosphere as discussed in the text.}\\
\multicolumn{12}{l}{$e$: Luminositity in the \ion{Ne}{ix} resonance line at 13.45 \AA~not considering the absorption.}\\
\multicolumn{12}{l}{$f$: Luminositity in the \ion{Ne}{ix} resonance line at 13.45 \AA~considering the absorption from the
stellar atmosphere as discussed in the text.}\\
\label{tab:Results}
\end{tabular}
\end{table*}

In all the simulations, during the first $100-200$ s, the system follows
the same evolution as described in Paper I. Figure \ref{fig:oscillations}
shows, as an example, the evolution of plasma temperature, density,
pressure and velocity for a stream with $n_{\rm acc}=10^{11}$ cm$^{-3}$,
$u_{\rm acc}=400$ km~s$^{-1}$ and $\zeta=1.0$. The accretion stream
penetrates the chromosphere and generates a transmitted shock and a
reverse shock. After this transient phase, the chromosphere stops the
flow penetration where the ram pressure of the post-shock plasma equals
the thermal pressure of the chromosphere. The reverse shock progressively
builds up a nearly isothermal slab at the post-shock temperature (heating
phase; left panels of Fig. \ref{fig:oscillations}). During this phase
the intensity of the radiative cooling at the base of the slab gradually
increases. The heating phase ends when the radiative cooling triggers
a thermal instability that robs the post-shock plasma of pressure support,
causing the material above the cooled layer to collapse back (cooling
phase; right panels of Fig. \ref{fig:oscillations}). Consequently the
reverse shock moves downwards to the chromosphere reducing the post-shock
zone thickness. After the hot slab has disappeared, a new slab is
re-built by the reverse shock, starting a new cycle of quasi-periodic
shock oscillations.  The post-shock zone is separated from the cold
chromosphere by a very steep transition region located where the
chromospheric thermal pressure equals the ram pressure of the post-shock
plasma. The thermal conduction drains energy from the shock-heated
plasma to the chromosphere through this transition region, acting as an
additional cooling mechanism (see also \citealt{Orlando2009A&A}). The
effects of the heat conduction are the largest in the simulations with
the highest post-shock temperatures (i.e. the highest stream velocities).

\begin{figure}[bt]
\includegraphics[width=8.5cm]{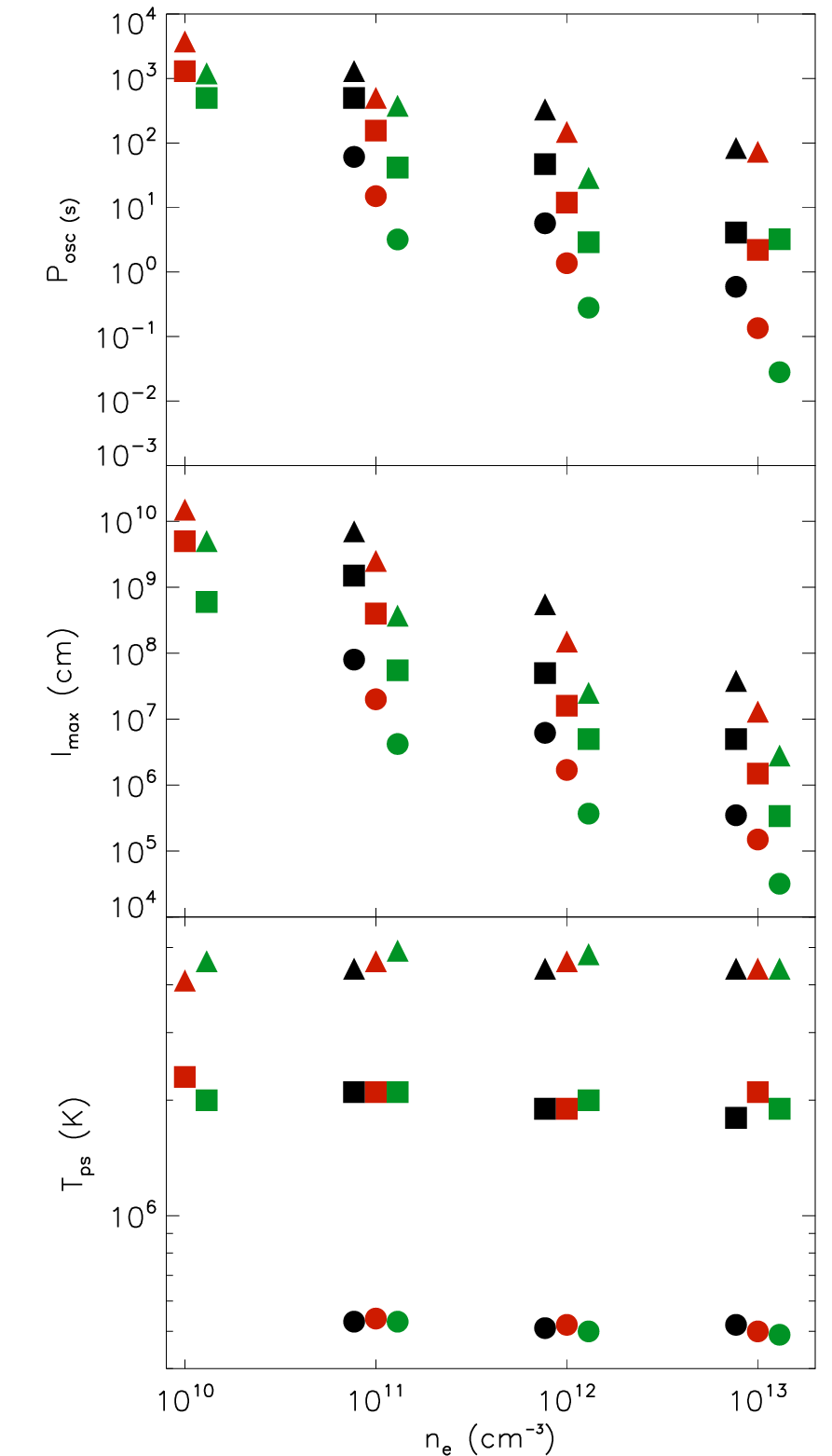}
\caption{Periods of the shock oscillations ($P_{\rm osc}$, upper
panel), maximum extension of the post-shock zone ($l_{\rm max}$, middle
panel) and the emission-measure-weighted temperature ($T_{\rm
ps}$, bottom panel)  of all the cases investigated. The labels on 
the x-axis of the bottom panel indicate the densities of accretion stream.  
Different symbols indicate different velocities:
$u_{\rm acc}=200$ (circle), 400 (squares), and 600 (triangles) km
s$^{-1}$. Different colors indicate different metal abundances: $\zeta =
0.2$ (black), 1.0 (red) and 5.0 (green).  Note that data points have 
been spread over each density value to prevent overlapping of the symbols.}
\label{fig:absval}
\end{figure}

Figure \ref{fig:absval} and Table \ref{tab:Results} show the main
physical parameters characterizing the structure of the post-shock zone
after the initial transient phase, namely the maximum extension of the
post-shock slab ($l_{\rm max}$), the oscillation period ($P_{\rm osc}$), and
the emission-measure-weighted temperature ($T_{\rm ps}$) of the post-shock
zone. In order to define $l_{\rm max}$, we consider all the
material hotter than $T=3\times 10^{5}$ K as shocked plasma. The
steepness of the transition region makes $l_{max}$ poorly sensitive
to this choice. The periods of the oscillations have been determined
through a Fourier analysis of the length of the post-shock zone as a
function of time. The temperature $T_{\rm ps}$ has been calculated from the
emission measure distribution EM($T$) (see Sect. \ref{sect:X-ray_syn}),
considering only bins with $\log T({\rm K})>5$.

The maximum extension of the post-shock zone ranges between a factor
0.3 and 1.7 of the length estimated using Eq. \ref{eqn:lslab}, whereas
the post-shock temperature ranges between a factor 0.7 and 1.0 of the
temperature estimated using Eq. \ref{eqn:posttemp}. The simulated
post-shock temperatures are slightly lower than those derived from
Eq. \ref{eqn:posttemp} because the former are averaged over the whole
post-shock zone, whereas the latter are estimated at the shock front, at
the maximum extension of the slab. In addition, the simulated post-shock
temperatures are averaged over several shock oscillations, including
both the heating and the cooling phase of the evolution.
During the former, the temperature is slightly higher than that derived
from Eq. \ref{eqn:posttemp}, because the plasma velocity in the reference
frame of the shock is higher than $u_{\rm acc}$; the opposite is true
during the cooling phase.

The post-shock temperature depends on the accretion stream velocity,
whereas both the oscillation period $P_{\rm osc}$ and the length of
the post-shock zone $l_{\rm max}$ depend on the whole set of model
parameters (velocity, density and metal abundance of the stream). Figure
\ref{fig:absval} and Table \ref{tab:Results} show that $P_{\rm osc}$
and $l_{\rm max}$ both span $5-6$ orders of magnitude for the
range of parameters explored here. In the complex magnetospheric
accretion scenario, the mass density and velocity of the flow could
vary along the accretion stream cross section (see for instance
\citealt{Romanova2004ApJ}). Consequently, in the case of $\beta \ll 1$,
an accretion stream cannot be described by a single 1-D model but has
to be considered as a bundle of independent fibrils, each described
in terms of a different 1-D model, and each independent on the others
(with different instability periods and random phases of the shock
oscillations) due to the strong magnetic field which prevents mass and
energy exchange across magnetic field lines. We expect therefore that
accretion streams consisting of many ($200-300$) different fibrils
with different instability periods and random phases would produce no
periodic variations of the X-ray emission from shocked plasma,
as recently found in the case of TW~Hya (\citealt{Drake2009ApJ}).

As discussed in Sect. \ref{sect:absorption}, the key parameters
for understanding the effects of the absorption by the stellar
chromosphere on the X-ray emission from shocked accreted plasma are
the thickness of the shocked slab $l_{\rm max}$ and the sinking of the
slab in the chromosphere to the position $h_{\rm sink}$ at which the
ram pressure of the post-shock plasma equals the thermal pressure of
the chromosphere. Figure \ref{fig:absval} shows that $l_{\rm max}$ is
anti-correlated with the density of the accretion stream and with the
metal abundance, and is correlated with the stream velocity (see also
Eq. \ref{eqn:lslab}). On the other hand, $h_{\rm sink}$ is correlated with
both the stream density and velocity, the ram pressure of the post-shock
plasma being $P_{\rm ram} \propto \rho_{\rm acc} u_{\rm acc}^2$ and the
thermal pressure of the chromosphere decreasing with height. Consequently,
the shocked slab from high density streams with high metal abundances
is expected to hardly emerge from the dense chromospheric layers,
where the absorption is strong, because the post-shock zone is rooted
deeply in the chromosphere ($h_{\rm sink}$ is large) and the 
slab is rather thin ($l_{\rm max}$ is small). On the other hand,
streams with high velocity form extended post-shock zones (being $l_{\rm
max}\propto u_{\rm acc}^4$) and the shocked slab may easily emerge above
the chromosphere even if it is deeply rooted in the chromosphere, the
ram pressure being proportional to the square of velocity.

\subsection{Distribution of emission measure versus temperature
\label{sec:emdis}}

\begin{figure}[ht]
\includegraphics[width=8.5cm]{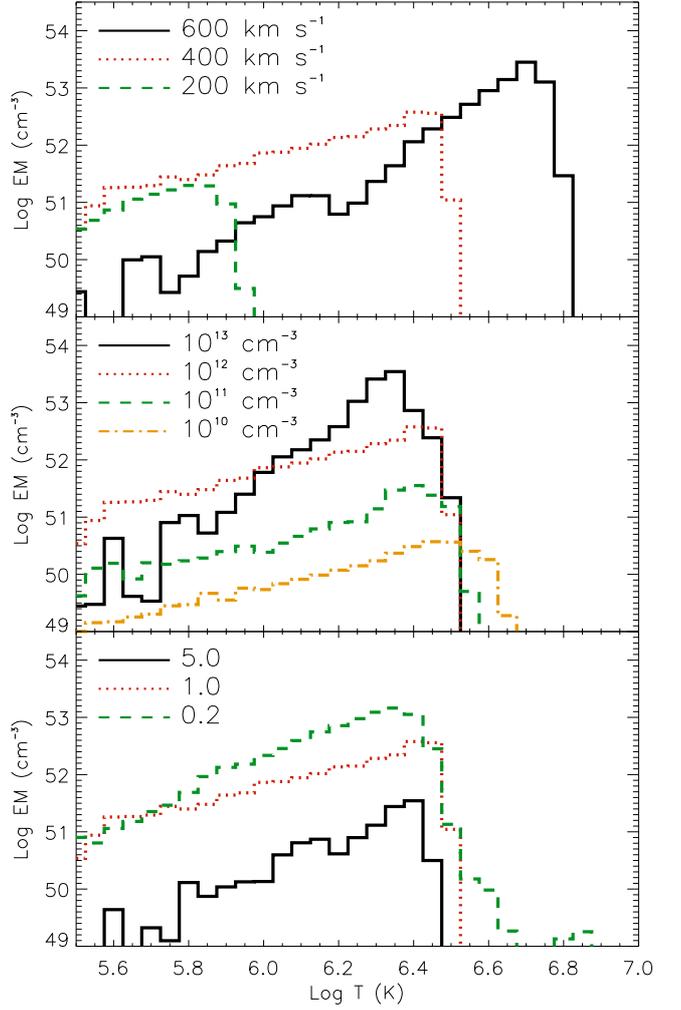}
\caption{Time-averaged emission measure distribution of the post-shock
plasma for accretion streams with the same metal abundance ($\zeta = 1.0$)
and density ($n_{\rm acc} = 10^{12}$ cm$^{-3}$) but different velocities
(upper panel), for streams with the same metal abundance ($\zeta = 1.0$)
and velocity ($u_{\rm acc} = 400$ km~s$^{-1}$) but different densities
(middle panel), and for streams with the same density ($n_{\rm acc} =
10^{12}$ cm$^{-3}$) and velocity ($u_{\rm acc} = 400$ km~s$^{-1}$) but
different metal abundances (lower panel).}
\label{fig:emdis}
\end{figure}

The distribution of emission measure versus temperature EM$(T)$
of the shock-heated plasma is a useful source of information of the
plasma components contributing to the X-ray emission and is directly
comparable with EM$(T)$ distributions derived from X-ray observations
(see, for instance, \citealt{Argiroffi2009A&A}). Figure \ref{fig:emdis}
shows how the time-averaged EM$(T)$ of the post-shock plasma varies as
a function of velocity (upper panel), density (middle panel), and metal
abundance (lower panel) of the accretion stream. The case with density
$n_{\rm acc}=10^{12}$ cm$^{-3}$, velocity $u_{\rm acc}=400$ km~s$^{-1}$
and solar abundance $\zeta=1$ (red dotted line in Fig.~\ref{fig:emdis})
is the reference in all the panels. In all cases, our model predicts a
monolithic distribution of emission measure of the post-shock plasma
that covers the entire range of values below the temperature of the
shock front. The ascending part of the EM$(T)$ distribution corresponds
to cooled post-shock plasma of the slab and to the transition region
between the shocked chromosphere and the hot slab. This characteristic
of the EM$(T)$ distribution cannot be reproduced by heuristic models
(e.g. \citealt{Lamzin1998ARep, Calvet1998ApJ, Argiroffi2007A&A})
and has been proved to be in agreement with observations
(e.g. \citealt{Argiroffi2009A&A}).

The total emission measure of plasma above 1 MK and the maximum
temperature of the emission measure distribution depend on the velocity
of the accretion stream (upper panel in Fig.~\ref{fig:emdis}). In fact,
the temperature of the post-shock plasma depends on the square of the
stream velocity (see Eq.~\ref{eqn:posttemp}), leading to a shift of the
EM($T$) profile towards higher temperature for higher value of $u_{\rm
acc}$. The thickness of the post-shock zone strongly increases with the
stream velocity (see Fig.~\ref{fig:absval} and Eq.~\ref{eqn:lslab}),
so that its volume and therefore the overall emission measure of the
shock-heated plasma increases with $u_{\rm acc}$.

Not surprisingly, the density of the stream determines the
overall emission measure in the hot slab (see middle panel in
Fig.~\ref{fig:emdis}): the higher the stream density, the higher the
emission measure of the slab. Note however that, since $l_{\rm acc}
\propto n_{\rm acc}^{-1}$ (see Eq. \ref{eqn:lslab}), the emission measure
of the post-shock zone depends only linearly on $n_{\rm acc}$ (not on
its square). Figure~\ref{fig:emdis} also shows that the ascending branch
of the EM$(T)$ distribution is steeper for larger values of density.

Finally, the EM$(T)$ distribution depends on the metal abundance:
the higher the value of $\zeta$, the lower the emission measure of
the slab. In fact, the length of the post-shock zone depends inversely
on the metal abundance (see Eq.~\ref{eqn:lslab}) as a consequence of
more efficient radiative cooling for larger $\zeta$. Consequently,
the emission measure of the hot slab decreases for increasing $\zeta$.

\subsection{X-ray emission \label{sec:high_energy}}

\begin{figure*}[ht]
\sidecaption
\includegraphics[width=12cm]{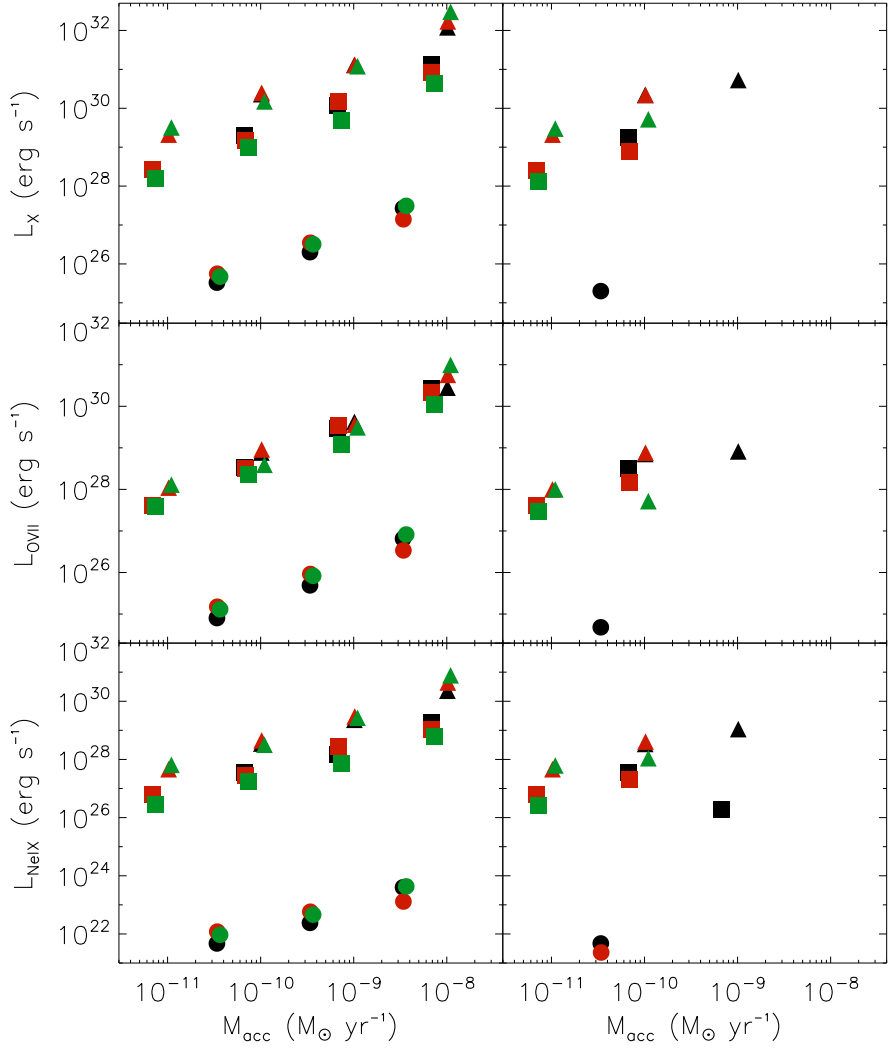}
\caption{Luminosities in the $[0.5-8.0]$ keV band (top panels) and in the
resonance lines of the \ion{O}{vii} (middle-panels) and \ion{Ne}{ix}
(bottom panels) as function of the mass accretion rates. Absorption
from the stellar chromosphere is considered only for the results reported
on right panels. Symbols and colors as in Fig. \ref{fig:absval}.}
\label{fig:lum_massrate} 
\end{figure*}

X-ray luminosities derived for each simulation are reported in
Table \ref{tab:Results} and in the Fig. \ref{fig:lum_massrate},
where they are plotted against the mass accretion rate derived
assuming the same accretion stream cross section ($A_{\rm str}=5\times
10^{20}$ cm$^2$) used for deriving emission measure distribution (see
Sect. \ref{sect:X-ray_syn}). The left panels show the results obtained
not considering absorption due to the stellar atmosphere, whereas right
panels show the results including the absorption effects using the method
discussed in Sect. \ref{sect:absorption}.

The dependence of the X-ray luminosities on the model parameters can
be easily explained by interpreting the results shown in the right
panels of Fig. \ref{fig:lum_massrate} in the light of the EM($T$)
distributions shown in Fig. \ref{fig:emdis} and of the relations between
the main properties of the post-shock zone and the stream parameters
in Fig. \ref{fig:absval}. The spread in the X-ray luminosity due to
the metal abundance is less than 0.5 dex even with the large range (a
factor 25) of metal abundances considered in this work. This result is
due to two different effects working in opposite directions. The EM($T$)
of the post-shock plasma is anti-correlated with the metal abundance
(bottom panel of Fig. \ref{fig:emdis}), but higher metal abundances
trigger higher emission in the soft X-ray band, because soft X-ray
emission is mainly due to line emission produced by heavy ions.

The X-ray luminosity is strongly correlated with the accretion stream
velocity: X-ray luminosities derived for streams at velocities $u_{\rm
acc}=400$ km~s$^{-1}$ and $u_{\rm acc}=600$ km~s$^{-1}$ differ for
about one order of magnitude, whereas X-ray luminosities produced by
streams with $u_{\rm acc}=200$ km~s$^{-1}$ are more than three orders
of magnitude lower than the others. In addition, the spread between the
streams at 400 and 600 km~s$^{-1}$ is much less in the \ion{O}{vii}
resonance line (left-central panel of Fig. \ref{fig:lum_massrate})
than in the $[0.5-8.0]$ keV band. Evidently, the relation between X-ray
luminosity and stream velocity strongly depends on the profile of the
EM($T$) distribution. In fact, low velocity (200 km~s$^{-1}$) accretion
streams cannot heat up plasma to temperatures greater than 1 MK (top
panel of Fig. \ref{fig:emdis}), so that the accretion energy is emitted
mainly in the extreme ultraviolet band and very faint X-ray emission
is expected. The peak of the EM($T$) for the stream at 400 km~s$^{-1}$
($\log T({\rm K})=6.4$) is very close to the peak of the emissivity
function of the \ion{O}{vii} resonance line and it is comparable to the
EM value at $\log T({\rm K})=6.4$ for the case at 600 km~s$^{-1}$,
explaining the small difference between the \ion{O}{vii} luminosities
for streams at 400 and 600 km s$^{-1}$.

\begin{figure}[ht]
\includegraphics[width=8.5cm]{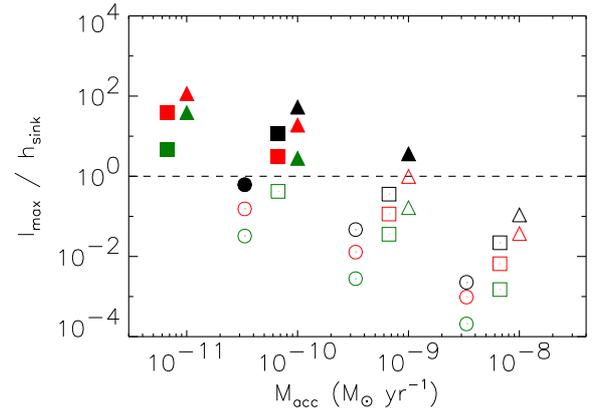}
\caption{The ratio $l_{max}/h_{sink}$ as function of the
mass accretion rate. Filled and empty symbols indicate if X-ray emission
from the stream is observable or absorbed, respectively.
Symbols and colors as in Fig. \ref{fig:absval}.}
\label{fig:stan-off_height}
\end{figure}

The comparison between the X-ray luminosities reported in the right and
left panels of Fig. \ref{fig:lum_massrate}, as well as the relation between
the ratio $l_{max}/h_{sink}$ and the model parameters (Fig.\ref{fig:stan-off_height}), 
allow us to highlight the role of absorption from the stellar atmosphere 
on the observability of the post-shock zone in the X-ray band as a 
function of the accretion stream properties. Specifically, we find that:

\begin{itemize}

\item Low density ($n_{\rm acc}\leq 10^{11}$ cm$^{-3}$) and high velocity
($u_{\rm acc}\geq 400$ km~s$^{-1}$) streams form shocked slabs with X-ray
emission poorly absorbed by the surrounding stellar chromosphere, the
thickness of the post-shock zone being much larger than the sinking of
the stream in the chromosphere ($l_{\rm max} \gg h_{\rm sink}$), so that
most of the emitting plasma is located in the shallower and low density
portion of the chromosphere.

\item High density ($n_{\rm acc}\geq 10^{13}$ cm$^{-3}$) and low velocity
($u_{\rm acc} \leq 200$ km~s$^{-1}$) accretion streams form shocked
slabs with X-ray emission strongly absorbed by the chromosphere, being
in these cases $l_{\rm ps} \ll h_{\rm sink}$.

\item The effects of the absorption in the case of intermediate density
($n_{\rm acc}\sim 10^{11}-10^{12}$ cm$^{-3}$) 
streams are less evident, especially for those cases with 
$l_{\rm max}\sim h_{\rm sink}$. Our results suggest that except for the stream 
with $u_{\rm acc}=600$ km~s$^{-1}$ and metal abundance $\zeta=0.2$, all 
the streams with density $n_{\rm acc}\approx 10^{12}$ cm$^{-3}$ generate 
shocked slab whose X-ray emission is strongly absorbed 
(see Fig. \ref{fig:stan-off_height}). However, these 
intermediate cases probably
require a more accurate analysis of the absorption effect that considers:
a) the absorption dependence on the wavelength; b) a detailed model
describing the absorption from the stellar atmosphere with a more smooth 
transition between the optically thin and the optically thick case and 
different angles of view; c) different metal abundances in the accretion stream and 
in the surrounding  atmosphere.

\end{itemize}

It is worth to emphasize that if accretion streams are not uniform
in density, the chromospheric absorption triggers a selection
effect, absorbing preferentially the X-ray emission from high
density plasma components. In addition, since the chromospheric
absorption does not equally affect the different density tracers
(e.g. \ion{O}{vii} and \ion{Ne}{ix} He-like triplets; see right panels
in Fig. \ref{fig:lum_massrate}), we expect that, in general, the results
of the use of these tracers may differ significantly (see also
discussion at the end of Sect. \ref{sec4}).

\section{Discussion}
\label{sec4}


Observational data and 3D magnetospheric models of the star-disk
interaction in young stars draw a very complex scenario
(see \citealt{Bouvier2007PPV} for a review of the recent results),
where circumstellar gas accretes onto the star through many streams
observed from different angles of view and, likely, 
non-uniform in density. Our modeling focuses on the
portion of the stream impacting on the chromosphere 
where a shock develops and assumes the stream with a
constant mass accretion rate. It is worth to emphasize, however
that the model results presented here can be combined together
in a tool-box like fashion, if the plasma-$\beta \ll 1$ (
assumption which is valid for many CTTSs as discussed in 
Sect. \ref{par:spacepar}), allowing us to build up inhomogeneous
streams as bundles of independent fibrils, each describable in
terms of a different 1-D model 
(see discussion in Sect. \ref{sec:post-shock}). This method 
can be easily applied to investigate the X-ray emission expected 
from a 3D density structured accretion flow, adding together the 
contribution to the emission of the distribution of fibrils each
with its own density. In the following, we apply this idea to very 
simplified cases. In the future the method could be applied to describe 
more complex configurations.

The hypothesis that the soft X-ray emission from CTTSs is mainly due
to post-shock accreting plasma is supported by the evidence that the
emission originates from high density plasma ($n_{\rm e}>10^{11}$
cm$^{-3}$). This evidence has been found in all (six) CTTSs with mass
lower than 1.2 $M_{\sun}$ observed to date by high resolution X-ray
spectrographs. In addition, no signatures of high density plasma have
been revealed in current X-ray observations of weak-line T Tauri stars
(e.g. \citealt{Kastner2004ApJ, Argiroffi2005A&A}) in which accretion is
known to be absent.

However, although the soft X-ray excess in CTTSs can be interpreted as
due to accretion shocks, some observational results are still not easily
explained in terms of post-shock accreting plasma: a) if we assume to
observe all the energy emitted by the post-shock zone, the accretion rates
derived from X-ray data are generally about $1-2$ orders of magnitude
lower than those derived from optical data (\citealt{Schmitt2005A&A,
Gunther2007A&A, Argiroffi2009A&A}; Curran et al. 2010, in prep.); b)
plasma densities derived from the \ion{O}{vii} triplet are generally lower
than those derived from the \ion{Ne}{ix} triplet \citep{Brickhouse2010ApJ},
although we expect the opposite result as the \ion{Ne}{ix} triplet forms
at higher temperatures than \ion{O}{vii} where accretion shock models
predict lower densities (e.g. \citealt{Sacco2008A&A}); c) no signatures
of high density plasma have been found in the high mass ($M> 2 M_{\sun}$)
accreting objects, namely T Tau, AB Aur and HD 163296 \citep{Gudel2007A&A,
Telleschi2007A&A, Gunther2009A&A}.

The above issues can be addressed in the light of the results discussed
in Sect. \ref{sec3}. The comparison of the right and left panels of
Fig. \ref{fig:lum_massrate} highlights the importance of the absorption
from the optically thick plasma of the chromosphere in the selection
of post-shock plasma components that produce observable emission in
the X-ray band. As a consequence, we found that the X-ray luminosity of
the post-shock zone as well as the mass accretion rate derived from the
emerging X-ray emission strongly depend on absorption effects.

Our exploration of the model parameter space has shown that only a
fraction of modeled streams predict post-shock plasma with measurable
X-ray emission. Interestingly, in five out of six CTTSs (MP Mus, BP
Tau, TW Hydrae, V4046 Sgr and RU Lupi) observed with high resolution
X-ray spectrographs, the pre-shock densities derived from the He-like
triplet ($10^{11}\la n_{\rm acc}\la 10^{12}$ cm$^{-3}$), the escape velocities
($400\la u_{\rm acc}\la 600$ km~s$^{-1}$) and the metal abundances
($\zeta\leq 1$) lie in the ranges of values that, according to the plots in
Fig. \ref{fig:lum_massrate}, are the most suitable for the observation
of X-ray emission from the post-shock accreting plasma. The CTTS Hen
3-600 lies at the boundaries of our range of observability due to its
low escape velocity ($u_{\rm acc}\sim290$ km~s$^{-1}$) and the nearly solar
metal abundances ($\zeta\sim 1.0$).

Our results can help us to reconcile the (apparent) discrepancy found
between the accretion rates measured from X-rays and optical data
(\citealt{Schmitt2005A&A, Gunther2007A&A, Argiroffi2009A&A}; Curran
et al. 2010, in prep.). In fact, in the case of accretion streams with
non-uniform mass density along the stream cross section (or in the case
of multiple streams with different density simultaneously present on the
star), we expect to observe X-rays preferentially from the low density
portion of the post-shock plasma, the high density shocked plasma
being strongly absorbed. For instance, as suggested by 3D MHD models of
the star-disk system \citep{Romanova2004ApJ}, accretion streams may
consist of a high density central region, surrounded by material at lower
densities. Consequently, the shocked slab should be constituted by a
dense core (rooted deeply in the chromosphere and with a short stand-off
height) surrounded by less dense shocked material (located shallower in
the chromosphere and with a large stand-off height). According to our
results in Sect. \ref{sec3}, the X-ray emission from the dense core is
expected to be largely absorbed by the stellar chromosphere and by the
accretion stream itself, whereas X-rays emitted from the (less dense)
shocked plasma at the boundary of the slab may suffer only a minor
absorption by the chromosphere. All the shocked plasma contributes to
the optical emission, whereas only a fraction of it contributes to the
X-ray emission. The result is that the accretion rates deduced from
optical observations are expected to be larger than the rate values
deduced from X-ray observations.

Under the assumption that the accretion streams are inhomogeneous and
taking into account the chromospheric absorption, our model predicts that
different He-like triplets, in general, measure different densities of
the emitting plasma. In fact, since the effects of absorption increases
with wavelength, \ion{Ne}{ix} (13.45 {\AA}) emission is expected to be
less absorbed than \ion{O}{vii} (21.60 {\AA}) emission. Consequently,
the high density plasma (i.e. the component suffering more absorption
effects) should contribute more to \ion{Ne}{ix} than to \ion{O}{vii}
emission. In the case of inhomogeneous streams, therefore the average
density of the plasma leading to \ion{Ne}{ix} emission is expected to
be larger that that of plasma leading to \ion{O}{vii} emission. This
argument can be checked with a simple case of an accretion stream with
velocity $u_{\rm acc}=600$ km~s$^{-1}$, metal abundance $\zeta=0.2$ and
two components with the same cross section, one with density $n_{\rm
acc}=10^{11}$ cm$^{-3}$ and the other with $n_{\rm acc}=10^{12}$
cm$^{-3}$. From Table \ref{tab:Results}, we obtain that the X-ray
luminosities in the \ion{O}{vii} resonance line due to the low and high
density stream components are $\log(L_{\rm OVII})=28.85$ and $\log(L_{\rm
OVII})=28.9$ erg~s$^{-1}$, respectively, whereas the luminosities in
the \ion{Ne}{ix} resonance line are $\log(L_{\rm NeIX})=28.53$ and
$\log(L_{\rm NeIX})=29.04$ erg~s$^{-1}$. In other words, the denser
stream contribute for 73\% to the \ion{Ne}{ix} emission, but only for
53\% to the \ion{O}{vii} emission. The average density measured with
the \ion{Ne}{ix} triplet is expected therefore to be larger than that
measured with \ion{O}{vii} triplet although \ion{Ne}{ix} lines form at
temperatures higher than \ion{O}{vii} lines.

Finally, our results do not allow us to formulate a unique hypotheses
to explain why soft X-ray emission from high density plasma have not been
detected in high mass young accreting objects (e.g. T Tau, AB Aur and HD
163296; \citealt{Gudel2007A&A, Telleschi2007A&A, Gunther2009A&A}). In
particular two hypotheses can be formulated: a) accretion streams in
these objects are, in general, characterized by densities higher than
$n_{\rm acc}=10^{12}$ cm$^{-3}$, so that the X-ray emission from shocked
material is fully absorbed by the chromosphere; b) accretion streams are
much less dense than streams of low-mass stars ($n_{\rm acc}<10^{10}$
cm$^{-3}$), so that the resulting X-ray emission from the shocked material
cannot be distinguished from the coronal emission. Further analysis of
optical and X-ray observations of high mass young accreting objects is
needed to assess if the lack of X-ray emission from high density plasma
is a general feature of these objects and if some hints on the density of
accretion streams can be obtained.

\section{Conclusions}
\label{sec5}

We performed an intensive simulation campaign, using the 1-D hydrodynamic
model introduced in Paper I, exploring the model parameter space which
is representative of almost all low-mass CTTSs observed to date in the
X-ray band. The model describes the impact of an accretion stream onto
the chromosphere of a CTTS under the hypothesis that $\beta \ll 1$.
Our aims include: 1) to investigate the physical properties of shocked
slab resulting from the stream impact as a function of the density,
velocity and metal abundance of the accretion stream and 2) to investigate
the observability of the post-shock plasma in the X-ray band, taking
into account the absorption from the optically thick plasma of the
surrounding chromosphere. Our main results can be summarized as follows:

\begin{enumerate}

\item The post-shock zone resulting from the impact of the accretion
stream onto the stellar surface is characterized by temperatures ranging
from 0.5 to 5 MK (depending on the stream velocity) and by stand-off
heights of the post-shock zone spanning six orders of magnitude from
$\approx 10^4$ to $\approx 10^{10}$ cm (mainly depending on the stream
density).

\item In all the case the post-shock zone oscillates
quasi-periodically. Oscillations are composed by a heating phase during
which the accretion shock builds up a slab of hot plasma and a cooling
phase during which the post-shock zone cools down under the effect of
thermal instabilities at its base. The oscillation period ranges from
$\approx 3\times 10^{-2}$ to $\approx 4\times10^{3}$ s within the space
of physical parameters. However, these oscillations are very difficult
to observe, because the accretion stream is, most likely, inhomogeneous
and clumped (i.e. with variable accretion rate) and, in the case of
$\beta \ll 1$, constituted by several different fibrils with different
instability periods and random phases, leading to no evident periodic
variations of the X-ray emission\footnote{In the case of $\beta \geq 1$,
see discussion in \cite{Orlando2009A&A}.}.

\item The effect of the absorption from the local chromosphere on
the X-ray emission from the post-shock zone strongly depends on the
accretion stream properties. The stream density is a key parameter to
make observable in the X-ray band the post-shock material: the higher
the stream density, the thinner and more deep rooted in the chromosphere
the shocked slab, and the larger the absorption of X-ray emission from
the slab. We found that, in general, high density accretion streams
($n_{\rm acc}>10^{12}$ cm$^{-3}$) produce shocked slab which are strongly
absorbed. Shocked plasma from high velocity streams is more easily
observable than that from low velocity ones, the stand-off heights of
the shocked slab increasing rapidly with the stream velocity.

\item Our results suggest that the discrepancy between the mass accretion
rates derived from optical and X-ray data as well as the different
densities measured, in general, from the \ion{O}{vii} and \ion{Ne}{ix}
line triplets diagnostics can be explained if accretion streams are
inhomogeneous (and/or multiple streams with different densities are
present simultaneously). In fact, only the light component of the
post-shock plasma is preferentially observed in the X-ray band, leading
to an underestimation of the mass accretion rate with respect to the
rate value deduced from optical observations. Also, the absorption in
\ion{O}{vii} lines is larger than in \ion{Ne}{ix} lines and, therefore,
the weight of dense plasma is expected to be larger in \ion{Ne}{ix}
than in \ion{O}{vii}.

\end{enumerate}

\begin{acknowledgements} Research on X-rays from young stars by G.S. 
is supported by NASA/Goddard XMM-Newton Guest Observer Facility grants
NNX09AT15G and NNX09AC11G and NASA Astrophysics Data Analysis program 
grant NNX09AC96G to RIT. This work was supported in part by the
Italian Ministry of University and Research (MIUR) and by Istituto
Nazionale di Astrofisica (INAF). We acknowledge support through the
EU Marie Curie Transfer of Knowledge program PHOENIX under contract
No. MTKD-CT-2005-029768. The software used in this work was in part
developed by the DOE-supported ASC / Alliance Center for Astrophysical
Thermonuclear Flashes at the University of Chicago. The simulations have
been executed at the HPC facility (SCAN) of the INAF - Osservatorio
Astronomico di Palermo.
\end{acknowledgements}

\bibliographystyle{aa} 
\bibliography{biblio}

\end{document}